\input{psfig}

\documentstyle[a4]{article}

\begin{document}

%\author{}
%\maketitle

\begin{center}
{\Large Controlling\ collapse in Bose-Einstein condensates by temporal
modulation of the scattering length} \bigskip

Fatkhulla Kh. Abdullaev$^{1,2}$, Jean Guy Caputo$^{3,4}$, Roberto A. 
Kraenkel$^{1}$, and Boris A. Malomed$^{1,5}$

\bigskip

$^1$ Instituto de Fisica Teorica -- UNESP, R. Pamplona 145, 01405-900 
S\~{a}o
Paulo, Brazil

$^{2}$ Physical-Technical Institute, Uzbek Academy of Sciences, 2-b
Mavlyanov str., 700084 Tashkent, Uzbekistan
\footnote{ permanent address}

$^3$ Laboratoire de Mathematiques INSA de Rouen, BP 8, 76131
Mont-Saint-Aignan cedex, France 
\footnote{ permanent address}

$^4$ Laboratoire de Physique th\'eorique et modelisation, \\
Universit\'e de Cergy-Pontoise and C.N.R.S.

$^5$Department of Interdisciplinary Studies, Faculty of Engineering, 
Tel
Aviv University, Tel Aviv 69978, Israel\footnote{ permanent address}

\bigskip

{\bf Abstract}
\end{center}

We consider, by means of the variational approximation (VA) and direct
numerical simulations of the Gross-Pitaevskii (GP) equation, the 
dynamics of 2D
and 3D condensates with a scattering length containing constant and
harmonically varying parts, which can be achieved with an ac magnetic 
field
tuned to the Feshbach resonance. For a rapid time modulation, we
develop an approach based on the direct averaging of the GP equation,
without using the VA. In the 2D case, both VA and direct simulations, 
as
well as the averaging method, reveal the existence of stable 
self-confined
condensates without an external trap, in agreement with
qualitatively similar results recently reported for spatial solitons in
nonlinear optics. In the 3D case, the VA again predicts the
existence of a stable self-confined condensate without a trap. 
In this case, direct simulations demonstrate that the stability 
is limited in time, eventually switching into collapse, even 
though the constant part of the scattering length
is positive (but not too large). Thus a spatially uniform ac magnetic
field, resonantly tuned to control the scattering length, may play the 
role
of an {\em effective trap} confining the condensate, and sometimes 
causing
its collapse.

\newpage

\section{Introduction}

Collisions between atoms play a crucially important role in the 
dynamics of
Bose-Einstein condensates (BECs). As is commonly known, the collisions 
are
accounted for by the cubic term in the corresponding Gross-Pitaevskii 
(GP)
equation that describes BEC dynamics in the
mean-field approximation. The coefficient in front of the cubic term,
proportional to the collision scattering length, may be both positive 
and
negative, which corresponds, respectively, to repulsive and attractive
interactions between the atoms \cite{Review}. In the case of an 
attractive
interaction, a soliton may be formed in an effectively one-dimensional 
(1D)
condensate \cite{soliton}; however, in 2D and 3D cases the attraction 
results in the collapse of the condensate ({\it weak} and {\it strong} 
collapse,
respectively \cite{Berge_Review}) if the number of atoms exceeds a 
critical
value \cite{Review}.

Recently developed experimental techniques \cite{JILA} make it possible 
to
effectively control the sign of the scattering length using an
external magnetic field because the interaction constant can be
changed through the
Feshbach resonance \cite{Kagan}. This technique makes it possible to 
quickly
reverse (in time) the sign of the interaction from repulsion to 
attraction,
which gives rise, via the onset of collapse, to an abrupt
shrinkage of the condensate, followed by a burst of emitted atoms and
the formation of a stable residual condensate \cite{JILA}.

A natural generalization of this approach for controlling the strength 
and
sign of the interaction between atoms and, thus, the coefficient in 
front of
the cubic term in the corresponding GP equation, is the application of 
a
magnetic field resonantly coupled to the atoms and consisting, in the
general case of dc and ac components. The dynamical behavior of 2D and 
3D 
condensates in this case is then an issue of straightforward physical 
interest, as it may be readily implemented in experiments. This is
the subject of the present work.

It is noteworthy that, in the 2D case, this issue is similar to a 
problem
which was recently considered in nonlinear optics for (2+1)D spatial
solitons (i.e., self-confined cylindrical light beams) propagating 
across
a nonlinear bulk medium with a layered structure, so that the size 
\cite
{berge} and, possibly, the sign \cite{towers} of the Kerr (nonlinear)
coefficient are subject to a periodic variation along the propagation
distance (it plays the role of the evolutional variable, instead of 
time, in
the description of optical spatial solitons). The same optical model 
makes
also sense in the (3+1)D case, because it applies to the propagation of 
``light
bullets'' (3D spatiotemporal solitons \cite{bullet}) through the 
layered
medium \cite{towers}. We will demonstrate below that the results 
obtained for
the BEC dynamics in the GP equation involving both a dc and ac
nonlinearity are indeed similar to findings reported in the framework 
of
the above-mentioned optical model. To the best of our knowledge, a GP
equation with a rapid time-periodic modulation of this type is proposed 
in
this work for the first time. Previously, a quasi-1D model was 
considered in
which the BEC stability was affected by a rapid temporal modulation 
applied to
the trapping potential (rather than to the spatially uniform 
nonlinearity
coefficient) \cite{CD} and the macroscopic quantum interference and 
resonances have been studied in \cite{Abd1}. Resonances in 2D and 3D 
BEC with periodically varying atomic scattering length has been considered in
\cite{Abd2,Adh}.

The paper is organized as follows. In section 2, we formulate the model 
to
be considered in this work and VA which will be employed to analyze the
model. In section 3, variational and numerical results are presented 
for the
2D case (the analysis based on VA also employs the Kapitsa averaging
procedure). Both approaches demonstrate the existence of a stable
self-sustained condensate, in a certain region of parameter space, 
so that the condensate can be effectively confined and maintained by 
means of
a spatially uniform resonant ac magnetic field, without any trapping
potential. In section 3, we also develop an alternative analytical 
approach,
based on the application of the averaging procedure directly to the GP
equation, without using the VA. Results produced by this approach
confirm those obtained by means of VA. In section 4, we show that the 
results
for the 3D case are essentially different from the ones in the 2D case. 
Here VA also
predicts the possibility of a stable condensate, while direct 
simulations
demonstrate that the stability is limited in time, finally giving way 
to
collapse; a noteworthy fact is that, while VA per se still provides
reasonable results in the 3D case, the averaging procedure, if combined 
with
VA, may yield completely wrong predictions in this case. A nontrivial
feature demonstrated by direct simulations in the 3D case is that the 
ac
component of the nonlinearity may give rise to collapse even in the 
case
when the dc (constant) component corresponds to repulsion. The paper is
concluded by section 5.

\section{The model and variational approximation}

We take the mean-field GP equation for the single-particle wave 
function in
its usual form, 
\begin{equation}
i\hbar \frac{\partial \psi }{\partial t}=-\frac{\hbar ^{2}}{2m}\Delta 
\psi 
%+ \frac{1}{2}m\Omega ^{2}r^{2}\psi 
+g|\psi |^{2}\psi 
% +V(r,t)\psi 
,  \label{GPE}
\end{equation}
where $\Delta $ is the 2D or 3D Laplacian, $r$ is the corresponding 
radial
variable
%, $\Omega ^{2}$ determines the strength of the trapping potential,
and $g=4\pi \hbar ^{2}a_{s}/m$, where $a_{s}$, $m$ are respectively 
the atomic scattering
length and mass. As indicated above we will assume the scattering 
length
to be time-modulated so that the nonlinear coefficient in Eq. 
(\ref{GPE}) takes the form $g=g_{0}+g_{1}\sin (\chi t)$, where $a_{0}$ 
and 
$a_{1}$ are the amplitudes of dc and ac parts, and $\chi $ \ is the
ac-modulation frequency.

Usually an external trapping potential is included to stabilize the 
condensate.
We have omitted it because it does not play an essential role.
This is also the case in some other situations, e.g., the formation
of a stable Skyrmion in a two-component condensate \cite{Skyrmion}. In
fact, we will demonstrate that the temporal modulation of the nonlinear
coefficient, combining the dc and ac parts as in Eq. (\ref{nls}) may, 
in a
certain sense, replace the trapping potential. Another caveat 
concerning the
present model is that the frequency of the ac drive must be chosen far
enough from resonance with any transition between the ground state of 
the
condensate and excited states, otherwise the mean-field description 
based on
the GP equation will not be adequate.

We now cast Eq. (\ref{GPE}) in a
normalized form by introducing a typical frequency $\Omega \sim 2g n_{0}/\hbar$,
where $n_{0}$ is the condensate density and rescale
the time and space variables as $t'=\Omega t$
$r'=r \sqrt{2 m\Omega / \hbar}$. This leads to the following equation
where the $'$ have been omitted

\begin{equation}
i\frac{\partial \psi }{\partial t}=-\left( \frac{\partial 
^{2}}{\partial
r^{2}}+\frac{D-1}{r}\frac{\partial }{\partial r}\right) \psi 
%+\frac{1}{2} r^{2}\psi 
-\left[ \lambda _{0}+\lambda _{1}\sin (\omega t)\right] |\psi
|^{2}\psi ,  \label{nls}
\end{equation}
in which it is implied that $\psi $ depends only on $t$ and $r$, $D=2$ 
or $3$
is the spatial dimension, $\lambda _{0,1}\equiv -g_{0,1}/\left( \Omega
\hbar \right) $, $\omega \equiv \chi /\Omega $.

Note that $\lambda _{0}>0$ and $\lambda _{0}<0$ in Eq. (\ref{nls})
correspond, respectively, to the self-focusing and self-defocusing
nonlinearity. Rescaling the field $\psi $, we will set $\left| \lambda
_{0}\right| \equiv 1$, so that $\lambda _{0}$ remains a sign-defining
parameter.

The next step is to apply the VA to Eq. (\ref{nls}). This approximation 
was
originally proposed \cite{And} and developed in nonlinear optics, first 
for
1D problems and, later for multi-dimensional models (see a recent 
review 
\cite{Progress}). A similar technique was elaborated for the
description of the multidimensional BEC dynamics based on the GP 
equation. 
\cite{varBEC}.

To apply VA in the present case, we notice that the Lagrangian density
generating Eq. (\ref{nls}) is 
\begin{equation}
{\cal L}(\psi)=\frac{i}{2}\left( \frac{\partial \psi }{\partial t}\psi 
^{\ast }- 
\frac{\partial \psi ^{\ast }}{\partial t}\psi \right) -\left| 
\frac{\partial
\psi }{\partial r}\right| ^{2} + \frac{1}{2}
\lambda (t)|\psi |^{4},  \label{density}
\end{equation}
where $\lambda (t)\equiv \lambda _{0}+\lambda _{1}\sin (\omega t)$, and 
the
asterisk stands for the complex conjugation. The variational {\it 
ansatz}
for the wave function of the condensate is chosen as the Gaussian 
\cite{And},
\begin{equation}
\psi_g(r,t)=A(t)\exp \left( -\frac{r^{2}}{2a^{2}(t)}+\frac{1}{2}
ib(t)\,r^{2}+i\delta (t)\right) ,  \label{ansatz}
\end{equation}
where $A,a,b$ and $\delta $ are, respectively, the amplitude, width, 
chirp
and overall phase, which are assumed to be real functions of time. We 
did
not include the degree of freedom related to the coordinate of the
condensate's center, as the trapping potential, although not explicitly
included into the model, is assumed to prevent the motion of the 
condensate
as a whole.

Following the standard procedure \cite{Progress}, we insert the ansatz 
into
the density (\ref{density}) and calculate the effective Lagrangian, 
\begin{equation}
L_{{\rm eff}}=C_{D}\int_{0}^{\infty }{\cal L}(\psi_g)\,r^{D-1}dr,  
\label{effective}
\end{equation}
where $C_{D}=2\pi $ or $4\pi $ in the 2D or 3D cases, respectively.
Finally the evolution equations for the time-dependent parameters of 
the
ansatz (\ref{ansatz}) are derived from $L_{{\rm eff}}$ using 
the corresponding Euler - Lagrange equations. Subsequent
analysis, as well as the results of direct numerical simulations, are 
presented
separately for the 2D and 3D cases in the two following sections.

\section{The two-dimensional case}

\subsection{Variational approximation}

In the 2D case, the calculation of the effective Lagrangian 
(\ref{effective}
) yields 
\begin{equation}
L_{{\rm eff}}^{{\rm (2D)}} 
=\pi (-\frac{1}{2}a^{4}A^{2}\dot{b}-a^{2}A^{2}\dot{\delta}
-A^{2} -a^{4}A^{2}b^{2}+\frac{1}{4}\lambda
(t)\,a^{2}A^{4}),  \label{L2D}
\end{equation}
where the overdot stands for the time derivative. The Euler-Lagrange
equations following from this Lagrangian yield, the
conservation of the number of atoms $N$ in the condensate, 
\begin{equation}
\pi A^{2}a^{2}\equiv N={\rm const},
\end{equation}
an expressions for the chirp and the width, 
$$\dot{a} =2ab, \ \dot{b} = \frac{2}{a^{4}} -2b^{2} -\frac{\lambda(t)N}{2\pi a^{4}},$$ 
and a closed-form evolution
equation for the width: 
\begin{equation}
\frac{d^{2}a}{dt^{2}}=\frac{2(2-\lambda (t)N/2\pi)}{a^{3}}\,,  \label{a2}
\end{equation}
which we rewrite as 
\begin{eqnarray}
\frac{d^{2}a}{dt^{2}} &=&\frac{-\Lambda +\epsilon \sin (\omega 
t)}{a^{3}},
\label{a2d} \\
\Lambda &\equiv &2\left( \lambda _{0}N/(2\pi)-2\right),\,\epsilon \equiv 
-\lambda
_{1}N/\pi.  \label{lambda}
\end{eqnarray}

In the absence of an ac component, $\epsilon =0$, Eq. (\ref{a2d}) 
conserves
the energy $E_{{\rm 2D}}=\left( \dot{a}^{2}-\Lambda a^{-2}\right) /2$.
Obviously, $E_{{\rm 2D}}\rightarrow -\infty $ as $a\rightarrow 0$, if 
$\Lambda >0$, and $E_{{\rm 2D}}\rightarrow +\infty $ as $a\rightarrow 
0$, if 
$\Lambda <0$. This means that, in the absence of the ac component, the 
2D
pulse is expected to collapse if $\Lambda >0$, and to spread out if 
$\Lambda
<0$. The case $\Lambda =0$ corresponds to the critical number of 
particles in the condensate (the so-called ``Townes soliton''). 
Note that a numerically exact value of the critical number is 
(in the present notation) $N=1.862$ \cite{Berge_Review}, while the 
variational equation (\ref{lambda}) yields $N=2$ (if $\lambda 
_{0}=+1$).

It is natural to specially consider the case when the ac component of 
the
nonlinear coefficient oscillates at a high frequency. In this case, Eq. 
(\ref
{a2d}) can be treated analytically by means of the Kapitsa averaging
method. To this end, we set $a(t)=\bar{a}+\delta a$, with $|\delta 
a|<<| 
\bar{a}|$, where $\bar{a}$ varies on a slow time scale and $\delta a$ 
is a
rapidly varying function with a zero mean value. After straightforward
manipulations, we derive the following equations for the slow and rapid
variables, 
\begin{eqnarray}
\frac{d^{2}}{dt^{2}}\bar{a} &=&-\Lambda (\bar{a}^{-3}+6\bar{a}
^{-5}\left\langle \delta a^{2}\right\rangle )-3\epsilon \,\left\langle
\delta a\,\sin (\omega t)\right\rangle \,\bar{a}^{-4},  \label{slow} \\
\frac{d^{2}}{dt^{2}}\delta a &=&3\,\delta a\Lambda 
\bar{a}^{-4}+\epsilon
\sin (\omega t)\,\bar{a}^{-3}.  \label{fast}
\end{eqnarray}
where $\left\langle ...\right\rangle $\ stands for averaging over the 
period 
$2\pi /\omega $. A solution to Eq. (\ref{fast}) is 
\begin{equation}
\delta a(t)=-\frac{\epsilon \sin (\omega t)}{\bar{a}^{3}\left( \omega 
^{2}+3 
\bar{a}^{-4}\Lambda \right) },  \label{delta_a}
\end{equation}
the substitution of which into Eq. (\ref{slow}) yields the final 
evolution
equation for the slow variable, 
\begin{equation}
\frac{d^{2}}{dt^{2}}\bar{a}=\bar{a}^{-3}\left[ -\Lambda -\frac{3\Lambda
\epsilon ^{2}}{(\omega ^{2}\bar{a}^{4}+3\Lambda 
)^{2}}+\frac{3}{2}\frac{
\epsilon ^{2}}{\omega ^{2}\bar{a}^{4}+3\Lambda }\right] \,.  
\label{aa2d}
\end{equation}

To examine whether collapse is enforced or inhibited by the ac 
component of
the nonlinearity, one may consider Eq. (\ref{aa2d}) in the limit 
$\bar{a}
\rightarrow 0$. In this limit, the equation reduces to 
\begin{equation}
\frac{d^{2}}{dt^{2}}\bar{a}=(-\Lambda +{\frac{\epsilon ^{2}}{6\Lambda 
}}) 
\bar{a}^{-3}.  \label{asymp2D}
\end{equation}
It immediately follows from Eq. (\ref{asymp2D}) that, if the amplitude 
of
the high-frequency ac component is large enough, $\epsilon 
^{2}>6\Lambda
^{2} $, the behavior of the condensate (in the limit of small 
$\bar{a}$) is
exactly opposite to that which would be expected in the presence of the 
dc
component only: in the case $\Lambda >0$, bounce should occur rather 
than
collapse, and vice versa in the case $\Lambda <0$.

On the other hand, in the limit of large $\bar{a}$, Eq. (\ref{aa2d}) 
takes
the asymptotic form $d^{2}\bar{a}/dt^{2}=-\Lambda \bar{a}^{-3}$, which 
shows
that the condensate remains self-confined in the case $\Lambda >0$ 
i.e., if
the number of atoms exceeds the critical value. This
consideration is relevant if $\bar{a}$ though being large remains 
smaller 
than the limit imposed by an external trapping potential, should it 
be added to the model.
Thus, these asymptotic results guarantee that Eq. (\ref{aa2d}) gives 
rise to
a stable behavior of the condensate, both the collapse and decay 
(spreading
out) being ruled out if 
\begin{equation}
\epsilon >\sqrt{6}\Lambda >0\,.  \label{stable2D}
\end{equation}
In the experiments with for example with $^{7}$Li with the critical number
$\sim 1500$ atoms if we have initially 1800 atoms (i.e. $N/2\pi =2.2 $) 
to stabilize  the condensate this means that the atomic 
scattering length for $\lambda_{0} =1$ should be harmonically modulated with the 
amplitude  $\epsilon = 0.98$. 
In fact, the conditions (\ref{stable2D}) ensure that the right-hand 
side of
Eq. (\ref{aa2d}) is positive for small $\bar{a}$ and negative for large 
$\bar{a}$. This implies that Eq. (\ref{aa2d}) must give rise to a
stable fixed point (FP). Indeed, when the conditions (\ref{stable2D}) 
hold, 
the right-hand side of Eq. (\ref{aa2d}) vanishes at exactly one FP, 
\begin{equation}\label{st}
\omega ^{2}\bar{a}^{4}=\frac{3\epsilon ^{2}}{4\Lambda }+\sqrt{3\left( 
\frac{
3\epsilon ^{4}}{16\Lambda ^{2}}-1\right) }-3\Lambda \,,  
\label{stableFP}
\end{equation}
which can be easily checked to be stable through the calculation of an
eigenfrequency of small oscillations around it.

Direct numerical simulations of Eq. (\ref{a2d}) produce results (not 
shown
here) which are in exact correspondence with those provided by the 
averaging
method, i.e., a stable state with $a(t)$ performing small oscillations
around the point (\ref{stableFP}). The 3D situation shows a drastic 
difference because this correspondence breaks down, as shown 
in the next section.

For the sake of comparison with the results obtained by means of an
alternative approach in the next subsection, we also need an 
approximate
form of Eq. (\ref{aa2d}) valid in the limit of small $\Lambda $ (i.e., 
when
the number of atoms in the condensate is close to the critical value) 
and
very large $\omega $: 
\begin{equation}
\frac{d^{2}}{dt^{2}}\bar{a}=-\frac{\Lambda 
}{\bar{a}^{3}}+\frac{3}{2}\frac{
\epsilon ^{2}}{\omega ^{2}\bar{a}^{7}}\,.  \label{smallLambda}
\end{equation}

To estimate the value of the amplitude of the high-frequency ac
component necessary to stop the collapse, we note that a characteristic
trap frequency is $\Omega \sim 100$ Hz. So, for a modulation frequency 
$\sim 3$ kHz, which may be regarded as a typical ``high modulation
frequency'', the dimensionless $\omega $ is $\sim 30$. If the initial
dimensionless number of atoms is, for example, $N/2\pi=2.2$ so that 
according to Eq.  (\ref{lambda}), $\Lambda =0.4$ (this corresponds 
to the $^{7}$Li condensate with $\approx$ 1800 atoms, the critical number 
being $\approx$ 1500), and the parameters  of modulation are $\lambda_{0}=1,
\lambda_{1} =2.3, \epsilon = 10$, then 
the stationary value of the condensate width found 
from (\ref{st}) is $a_{st} = 0.8 l$,where   $l=\sqrt{m\Omega/\hbar}$
is the healing length.

Thus our analytical approach, based on the VA and the subsequent use of 
the
assumption that the number of atoms slightly exceeds the critical 
value,
leads to an important prediction: in the 2D case, the ac component of 
the
nonlinearity, acting jointly with the dc one corresponding to 
attraction,
may give rise not to collapse, but rather to a stable soliton-like
oscillatory condensate state which confines itself without the trapping
potential. It is relevant to mention that a qualitatively similar 
result,
viz., the existence of stable periodically oscillating spatial 
cylindrical
solitons in a bulk nonlinear-optical medium consisting of alternating 
layers
with opposite signs of the Kerr coefficient, was reported in Ref. \cite
{towers}, where this result was obtained in a completely analytical 
form on
the basis of the VA, and was confirmed by direct numerical simulations.

\subsection{Averaging of the Gross-Pitaevskii equation and Hamiltonian}

In the case of a high-frequency modulation, there is a possibility to
apply the averaging method directly to the 2D equation (\ref{nls}), 
without
using the VA. Note that direct averaging was applied to the 2D 
nonlinear
Schr\"{o}dinger equation (NLS) with a potential rapidly varying in 
{\it space}, rather than in time, in Ref. \cite{Kiv}, 
where the main results were
a renormalization of the parameters of the 2D NLS equation and a shift 
of the
collapse threshold. As we will see below, a rapid temporal modulation 
of
the nonlinear term in the GP equation leads to new effects, which do 
not
reduce to a renormalization. Namely, new nonlinear-dispersive and
higher-order nonlinear terms will appear in the corresponding effective 
NLS
equation [see Eq. (\ref{higher_order}) below]. These terms essentially
affect the dynamics of the collapsing condensate.

Assuming that the ac frequency $\omega $ is large, we rewrite the 2D
equation (\ref{nls}) in a more general form, 
\begin{equation}
i\partial \psi /\partial t+\Delta \psi +\lambda (\omega t)|\psi 
|^{2}\psi
=0,\   \label{omega}
\end{equation}
where $\Delta $ is the 2D Laplacian. To derive an equation governing 
the slow
variations of the field, we use the multiscale approach, writing 
the solution as an expansion in powers of $1/\omega $ and introducing 
the
slow temporal variables, $T_{k}\equiv \omega ^{-k}t,\ k=0,1,2...\,$, 
while
the fast time is $\zeta \equiv \omega t$. Thus, the solution is sought 
for
as 
\begin{equation}
\psi (r,t)=A(r,T_{k})+\omega ^{-1}u_{1}(\zeta ,A)+\omega 
^{-2}u_{2}(\zeta
,A)+...,  \label{psi}
\end{equation}
with $\left\langle u_{k}\right\rangle =0$, where $\left\langle
...\right\rangle $ stands for the average over the period of the rapid
modulation, and we assume that $\lambda _{0}=+1$ (i.e., the dc part of 
the
nonlinear coefficient corresponds to attraction between the atoms).

Following a procedure developed, for a similar problem, in Ref. 
\cite{Kath},
we first find the first and second corrections,{\bf \ } 
\begin{equation}
u_{1}=-i[\mu _{1}-\left\langle \mu _{1}\right\rangle ]|A|^{2}A,\,\mu
_{1}\equiv \int_{0}^{\zeta }\left[ \lambda (\tau )-\left\langle \lambda
_{1}\right\rangle \right] d\tau ,  \label{corr1}
\end{equation}
\begin{eqnarray*}
u_{2} &=&[\mu _{2}-\left\langle \mu _{2}\right\rangle
][2i|A|^{2}A_{t}+iA^{2}A_{t}^{\ast }+\Delta (|A|^{2}A)]- \\
&&|A|^{4}A[\frac{1}{2}[\left( \mu _{1}-\left\langle \mu 
_{1}\right\rangle
\right) ^{2}-2M]+\left\langle \lambda \right\rangle \left( \mu
_{2}-\left\langle \mu _{2}\right\rangle \right) ].
\end{eqnarray*}
Here$\mu_{2} = \int_{0}^{\zeta}(\mu_{1} -<\mu_{1} >)ds, M = 
(1/2)(<\mu_{1}^{2}> 
- <\mu_{1}>^{2})= (1/2)(<\lambda^{2}> - 1)$ (recall we have set 
$|\lambda_{0}| =1$).
 Using these
results, we obtain the following evolution equation for the
slowly varying field $A(x,T_{0})$, derived at the order $\omega ^{-2}$: 
\[
i\frac{\partial A}{\partial t}+\Delta A+|A|^{2}A+2M\left( 
\frac{\epsilon }{
\omega }\right) ^{2}[|A|^{6}A- 
\]
\begin{equation}
3|A|^{4}\Delta A+2|A|^{2}\Delta (|A|^{2}A)+A^{2}\Delta (|A|^{2}A^{\ast 
}]=0,
\label{higher_order}
\end{equation}
where $\epsilon $ is the same amplitude of the ac component as in Eq. 
(\ref
{lambda}). We
stress that Eq. (\ref{higher_order}) is valid in both 2D and 3D cases. 
In either case, it can be represented in the quasi-Hamiltonian form, 
\begin{equation}
\left[ 1+6M\left( \frac{\epsilon }{\omega }\right) ^{2}\left|
A\right| ^{4}\right] \frac{\partial A}{\partial t} =-i\frac{\delta 
H_{q}}{
\delta A^{\ast }},  \label{quasi} 
\end{equation}
\begin{equation}
H_{q} =\int dV\left[ \left| \nabla A\right| ^{2}-2M\left( \frac{
\epsilon }{\omega }\right) ^{2}\left| A\right| ^{8}-\frac{1}{2}\left|
A\right| ^{4}+ 4M\left( \frac{\epsilon }{\omega }\right) ^{2}\left| 
\nabla
\left( \left| A\right| ^{2}A\right) \right| ^{2}\right] ,  \label{H}
\end{equation}
where $dV$ is the infinitesimal volume in the 2D or 3D space. To cast 
this
result in a canonical Hamiltonian representation, one needs to properly
define the corresponding symplectic structure (Poisson's brackets), 
which is
not our aim here. However, we notice that, as it immediately follows 
from
Eq. (\ref{quasi}) and the reality of the (quasi-)Hamiltonian (\ref{H}), 
$H_{q}$ is an integral of motion, i.e., $dH_{q}/dt=0$.

For a further analysis of the 2D case, we apply a modulation theory
developed in Ref. \cite{Fibich}. According to this theory, the solution 
is
searched for in the form of a modulated Townes soliton. The
(above-mentioned) Townes soliton is a solution to the 2D NLS equation 
in the
form $\psi (r,t)=e^{it}R_{T}(r)$, where the function $R_{T}(r)$ 
satisfies
the boundary value problem 
\begin{equation}\label{BVP}
R_{T}^{\prime \prime }+r^{-1}R_{T}^{\prime }-R_{T}+R_{T}^{3}=0,~~
R_{T}^{\prime }(0)=0,~~R_{T}(\infty )=0. 
\end{equation}
For this solution, the norm $N$ and the Hamiltonian $H$ take the 
well-known
values, 
\begin{equation}
N_{T}\equiv \int_{0}^{\infty }R_{T}^{2}(r)rdr=N_{c}\equiv 1.862,~~
H_{T}=\int_{0}^{\infty }\left[ \left( R_{T}^{\prime }\right) 
^{2}-\frac{1}{2}
R_{T}^{4}(r)\right] rdr=0.  \label{Ncrit}
\end{equation}

The averaged variational equation (\ref{higher_order}) 
indicates an increase of the critical number of atoms for the
collapse, as opposed to the classical value (\ref{Ncrit}). Using the
relation (\ref{psi}), we find
\[
N_{{\rm crit}}=\int_{0}^{\infty }|\psi |^{2}rdr=N_{T}+ 
2M(\frac{\epsilon }{
\omega })^{2}I_{6}, 
\]
where $I_{6} = 11.178$(see Appendix 1).
This increase in the critical number of atoms is similar to the 
well-known energy enhancement of dispersion-managed solitons in 
optical fibers with periodically modulated dispersion \cite{DM,ac98}.

Another nontrivial perturbative effect is the appearance of a nonzero 
value
of the phase {\it chirp} inside the stationary soliton. We define the 
mean
value of the chirp as 
\[
b=\frac{\int_{0}^{\infty }{\rm Im}\left( \frac{\partial \psi }{\partial 
r}
\psi ^{\ast }\right) rdr}{\int_{0}^{\infty }r^{2}dr|\psi |^{2}}. 
\]
Making use of the expression (\ref{corr1}) for the first correction, we 
find 
\[
b=-\frac{\epsilon }{\omega }BM\left( \mu _{1}-\left\langle \mu
_{1}\right\rangle \right) ,\ B\equiv 3\frac{\int_{0}^{+\infty
}rdrR^{2}(R^{\prime })^{2}-(0.25)\int_{0}^{+\infty }drR^{4}}{
\int_{0}^{+\infty }r^{2}drR^{2}}=0.596. 
\]

To develop a general analysis, we assume that the solution with the 
number of
atoms close to the critical value may be approximated as a modulated 
Townes
soliton, i.e. 
\begin{equation}
A(r,t)\approx \left[ a(t)\right] ^{-1}R_{T}(r/a(t))e^{iS},\ S=\sigma 
(t)+
\frac{\dot{a}r^{2}}{4a},\,\dot{\sigma}=a^{-2}  \label{A}
\end{equation}
with some function $a(t)$ (where the overdot stands for $d/dt$). If the
initial power is close to the critical value, i.e., when $|N-N_{c}| << N_{c}$ 
and the perturbation is
conservative,i.e.
$$\mbox{Im}\int dV [A^{\ast}F(A)] =0$$
as in our case,  a method worked out in Ref. \cite{Fibich} makes it 
possible to
derive an evolution equation for the function $a(t)$, starting from the
approximation (\ref{A}). The equation of modulation theory for width is 
\begin{equation}
a^{3}a_{tt} = -\beta_{0} +\frac{\epsilon^{2}}{4M_{0}\omega^{2}}f_{1}(t),
\end{equation}
where $$\beta_{0} = \beta(0) - \frac{\epsilon^{2}f_{1}(0)}{(4M_{0}\omega^{2}},
\ \beta(0) = \frac{(N-N_{c})}{M_{0}},$$ and $M_{0}\equiv (1/4)\int_{0}^{\infty }r^{3}drR_{T}^{2}
\approx 0.55$. 
The auxiliary function is given by
\begin{equation}
f_{1}(t) = 2a(t)\mbox{Re}[\frac{1}{2\pi}\int dxdy F(A_{T})e^{-iS}(R_{T} + 
\rho\nabla R_{T}(\rho))].
\end{equation}
In the lowest-order approximation, the 
equation
takes the form (for the harmonic modulation) 
\begin{equation}
\frac{d^{2}a}{dt^{2}}=-\frac{\Lambda_{1}}{a^{3}}+\frac{C\epsilon 
^{2}}{\omega
^{2}a^{7}},  \label{smallLambda2}
\end{equation}
where $\Lambda_{1} = (N-N_{c})/M_{0} - C\epsilon^{2}/(\omega^{2}a_{0}^4)$
and $C$ is
\begin{equation}
C\equiv \frac{3}{M_{0}}\int_{0}^{\infty }d\rho [ 2\rho
R_{T}^{4}(R_{T}^{\prime })^{2}- \rho^{2}R_{T}^{3}(R_{T}^{\prime })^{3}-
\frac{1}{8}\rho R_{T}^{8}] \approx 39. 
\end{equation}
 Values of integrals are given in the Appendix 1.
Thus the averaged equation predicts the {\it arrest}  of collapse
by the rapid modulations  of the nonlinear term in the 2D GP equation.
The
comparison of Eq. (\ref{smallLambda2}) with its counterpart (\ref
{smallLambda}), which was derived by means of averaging the 
VA-generated
equation (\ref{a2d}), shows that both approaches lead to the same 
behavior near the collapse threshold. The numerical coefficients in the 
second 
terms are different due to the different  profiles of the Gaussian 
and Townes soliton.
 In this connection, it is relevant to mention a recent work 
\cite{comparison}, which has demonstrated that, generally, one may 
indeed
expect good agreement between results for 2D solitons produced by VA 
and by
the method based on the modulated Townes soliton.

 Let us estimate the value of the fixed point for the numerical 
simulations
performed in Ref.\cite{berge}. In this work the stable propagation of 
 soliton has been 
observed
for two step modulation of the nonlinear coefficient in 2D NLSE. The 
modulation of the nonlinear coefficient was 
$\lambda = 1 + \epsilon $ if $T>t>0$, and $\lambda = 1 - \epsilon$ 
for $2T > t > T$.
The parameters in the numerical simulations has been taken as $T =\epsilon =0.1, N/(2\pi) =11.726/(2\pi)$, 
with the 
critical number as
$N_{c} = 11.68/(2\pi)$. The map strength M is $M = \epsilon^2 T^{2}/24$. For this 
values we have
$a_{c} = 0.49$, that agreed with the value $a_{c} \approx 0.56$ following from 
the numerical experiment. 

Instead of averaging Eq. (\ref{nls}), one can apply the averaging 
procedure,
also based on the representation (\ref{psi}) for the wave function, 
directly
to the Hamiltonian of Eq. (\ref{nls}). As a result, the averaged 
Hamiltonian
is found in the form 
\begin{equation}
\bar{H}=\int dxdy[|\nabla A|^{2}+2M(\frac{\epsilon }{\omega 
})^{2}|\nabla
(|A|^{2}A)|^{2}-\frac{1}{2}|A|^{4}-6M(\frac{\epsilon }{\omega 
})^{2}|A|^{8}].
\label{averH}
\end{equation}
A possibility to stop the collapse, in the presence of a rapid periodic
modulation of the atomic scattering length, can be explained on the 
basis
of this Hamiltonian. To this end, following the pattern of the usual 
virial
estimates \cite{Berge_Review}, we note that, if a given field 
configuration
has compressed itself to a spot with a size $\rho $, where the 
amplitude of
the $A$-field is $\,\sim \aleph $, the conservation of the number of
particles $N$ [which may be applied to the $A$-field through the 
relation 
(\ref{psi})] yields the relation 
\begin{equation}
\aleph ^{2}\rho ^{D}\sim N  \label{aleph}
\end{equation}
(recall $D$ is the space dimension). On the other hand, the same 
estimate
for the strongest collapse-driving and collapse-stopping terms [the 
fourth and
second terms, respectively, in expression (\ref{averH})] $H_{-}$ and 
$H_{+}$ in the Hamiltonian yields 
\begin{equation}
H_{-}\sim -\left( \frac{\epsilon }{\omega }\right) ^{2}\aleph ^{8}\rho
^{D},\,~~H_{+}\sim \left( \frac{\epsilon }{\omega }\right) ^{2}\aleph 
^{6}\rho
^{D-2}\,.  \label{H-+}
\end{equation}
Eliminating the amplitude from Eqs. (\ref{H-+}) by means of the 
relation 
(\ref{aleph}), we conclude that, in the case of the catastrophic
self-compression of the field in the 2D space, $\rho \rightarrow 0$, 
both
terms $H_{\mp }$ take the same asymptotic form, $\rho ^{-6}$, hence the
collapse may be stopped, depending on details of the configuration.
However, in the 3D case the collapse-driving term diverges as $\rho 
^{-9}$,
while the collapse-stopping term has the asymptotic form $\,\sim 
\rho ^{-8}$, for $\rho \rightarrow 0$, hence in this case the collapse, 
generally speaking, {\em cannot} be prevented.

Lastly, it is relevant to mention that, although the quasi-Hamiltonian 
(\ref
{H}) is not identical to the averaged Hamiltonian (\ref{averH}), the 
virial
estimate applied to $H_{q}$ yields exactly the same result: the 
collapse can
be stopped in the 2D but not in the 3D situation. 

\subsection{Direct numerical results}

The existence of stable self-confined soliton-like oscillating 
condensate
states, predicted above by means of analytical approximations for the 
case 
(\ref{stable2D}), when the dc part of the nonlinearity corresponds to
attraction between the atoms, and the amplitude of the ac component is 
not
too small, must be checked against direct simulations of the 2D 
equation 
(\ref{nls}). In fact, it was quite easy to confirm this prediction [in 
the
case $\lambda _{0}=-1$, i.e., when the dc component of the nonlinearity
corresponds to repulsion, the direct simulations always show a decay
(spreading out) of the condensate, which also agrees with the above
predictions].

A typical example of the formation of a self-confined condensate,
supported by the combination of the self-focusing dc and sufficiently 
strong
ac components of the nonlinearity in the absence of an external trap, 
is
displayed in Fig. 1. On the left panel we show the pulse collapse at 
$t \approx 0.3$ in the absence of modulation. In the presence of 
modulation
the pulse is stabilized for about $40$ periods after which it decays.
Note the presence of radiation as the pulse adjusts to the modulation.

%\begin{figure}[ht] 
%\centerline{\psfig{figure=f01.ps,height=8cm,width=7cm,angle=-90}
%\psfig{figure=f02.ps,height=8cm,width=7cm,angle=-90}}
%%was \label{f07}
%\caption{A typical example of the formation of a self-confined
%condensate, revealed from direct simulations of Eq. (\ref{nls}) in the
%two-dimensional case. The left panel shows pulse collapse in the 
%absence
%of modulation for $t\approx 0.3$. The right panel shows the modulated 
%pulse with the same initial condition for $t\approx 0.6$.
%The parameters are $\lambda_0=2.4,~\lambda_1=0.85$
%$\omega = 100 \pi$ and $N=5$.}
%\label{Fig1}
%\end{figure}

\section{The three-dimensional case}

\subsection{The variational approximation and averaging}

The calculation of the effective Lagrangian (\ref{effective}) in the 3D 
case
yields 
\begin{equation}
L_{{\rm eff}}^{{\rm (3D)}}=\frac{1}{2}\pi ^{3/2}A^{2}a^{3}\left[ 
-{\frac{3}{2}}\dot{b}
a^{2}-2\dot{\delta}+\frac{1}{2\sqrt{2}}\lambda (t)\,A^{2}-\frac{3}{a^{2}}
-3b^{2}a^{2}\right] \,,  \label{L3D}
\end{equation}
cf. Eq. (\ref{L2D}). The Euler-Lagrange equations applied to this 
Lagrangian
yield the mass conservation, 
\[
\pi ^{3/2}A^{2}a^{3}\equiv N={\rm const}, 
\]
an expressions for the chirp, $$\dot{a} = 2ab,\ \dot{b} = \frac{2}{a^4} -2b^2 -
\frac{\lambda(t)N}{2\sqrt{2}\pi^{3/2}a^5}, $$ and the 
evolution
equation for the width of the condensate,
\begin{equation}
\frac{d^{2}a}{dt^{2}}=\frac{4}{a^{3}}-\frac{\lambda (t)}{\sqrt{2}\pi 
^{3/2}}\frac{N}{
a^{4}}\,.  \label{a3}
\end{equation}
Note the difference of Eq. (\ref{a3}) from its 2D counterpart 
(\ref{a2}).

As in the 2D case, we renormalize the amplitudes of the dc and ac 
components of
the nonlinearity, $\Lambda \equiv 2^{-1/2}\pi ^{-3/2}\lambda _{0}N$ and 
$\epsilon
\equiv -2^{-1/2}\pi ^{-3/2}\lambda _{1}N$, and cast Eq. (\ref{a3}) in the 
normalized
form, 
\begin{equation}
\frac{d^{2}a}{dt^{2}}=\frac{4}{a^{3}}+\frac{-\Lambda +\epsilon \sin 
(\omega
t)}{a^{4}}\,.  \label{a3d}
\end{equation}
In the absence of the ac term, $\epsilon =0$, Eq. (\ref{a3d}) conserves 
the
energy 
\[
E_{{\rm 3D}}=\frac{1}{2}\dot{a}^{2}+2a^{-2}-\frac{1}{3}\Lambda a^{-3}. 
\]
Obviously, $E_{{\rm 3D}}\rightarrow -\infty $ as $a\rightarrow 0$, if 
$\Lambda >0$, and $E_{{\rm 3D}}\rightarrow +\infty $ if $\Lambda <0$, 
hence
one will have collapse or decay (spreading out) of the pulse, 
respectively,
in these two cases.

Prior to applying the averaging procedure (as it was done above in the 
2D
case), we solved Eq. (\ref{a3d}) numerically, without averaging, to 
show
that (within the framework of VA) there is a region in parameter space 
where the
condensate, that would decay under the action of the repulsive dc
nonlinearity ($\Lambda <0$), may be stabilized by the ac component of 
the
nonlinearity, provided that its amplitude is sufficiently large. To 
this
end, we employed the variable-step ordinary differential equation (ODE) 
solver DOPRI5 \cite{hairer}, which is
a combination of the Runge-Kutta algorithm of the fourth and fifth 
orders,
so that the instantaneous truncation error can be controlled.

In Fig. 2 we show the dynamical behavior of solutions to Eq. 
(\ref{a3d}), in
terms of the Poincar\'{e} section in the plane $(a,\dot{a})$, obtained 
for 
$\Lambda =-1,\epsilon =100,\omega =10^{4}\cdot \pi $, and initial 
conditions 
$a(t=0)=0.3$, $0.2$, or $0.13$ and $\dot{a}(t=0)=0$. As it is obvious 
from
Fig. 2, in all these cases the solution remains bounded and the 
condensate
does not collapse or decay, its width performing quasi-periodic 
oscillations.

%\begin{figure}[ht]
%\centerline{\psfig{figure=f03.ps,height=6cm,width=8cm,angle=-90}
%}
%\caption{The Poincar\'{e} section in the plane $(a,\dot{a})$ for 
%$\Lambda=-1, \protect\epsilon=100, \protect\omega=10^4\dot\protect\pi 
%$,
%generated by the numerical solution of the variational equation 
%(\ref{a3d})
%with different initial conditions (see the text).}
%\label{Fig2}
%\end{figure}

In fact, the corresponding stability region in the parameter plane 
$(\omega
/\pi ,\epsilon )$ is small, see Fig. 3. It is also seen from Fig. 3 
that the
frequency and amplitude of the ac component need to be large to yield 
this
stability. Notice that, for frequencies larger than $10^{6}\cdot \pi $, 
the
width of the condensate $a(t)$ assumes very small values in the course 
of the
evolution (as predicted by VA) so that collapse may occur in practice
for the solution of the full equation (\ref{nls}). 

The stability is predicted by VA only for $\Lambda <0$, i.e., for a
repulsive dc component of the nonlinearity. In the opposite case,
the VA predict solely collapse.

%\begin{figure}[ht]
%\centerline{\psfig{figure=f04.ps,height=6cm,width=8cm,angle=-90}}
%\caption{The region in the $(\protect\epsilon, 
%\protect\omega/\protect\pi)$
%parameter plane where the numerical solution of Eq. (\ref{a3d}) with 
%$\Lambda=-1$ predicts stable quasiperiodic solutions in the 3D case. 
%Crosses
%mark points where stable solutions were actually obtained. Stars 
%correspond
%to the minimum values of the ac-component's amplitude 
%$\protect\epsilon$
%eventually leading to collapse of the solution of the full partial
%differential equation (\ref{nls}) with $\Lambda=-1$, see below.}
%\label{Fig3}
%\end{figure}

As $\omega $ is large enough in the stability region shown in Fig. 3, 
it
seems natural to apply the Kapitsa's averaging method to this case too.
Doing it the same way as was described in detail in the previous 
section for
the 2D case, we find the rapidly oscillating correction $\delta a(t)$, 
cf.
Eq. (\ref{delta_a}), 
\begin{equation}
\delta a=-\frac{\epsilon \sin (\omega t)\bar{a}}{\omega 
^{2}\bar{a}^{5}-12 
\bar{a}+4\Lambda },  \label{delta_a3D}
\end{equation}
and then arrive at the evolution equation for the slow variable 
$\bar{a}(t)$
[cf. Eq. (\ref{aa2d})]: 
\begin{equation}
\frac{d^{2}\bar{a}}{dt^{2}}=\bar{a}^{-4}\left[ 4\bar{a}-\Lambda +\frac{
2\epsilon ^{2}}{\omega ^{2}\bar{a}^{5}-12\bar{a}+4\Lambda }+\epsilon 
^{2} 
\frac{6\bar{a}-5\Lambda }{(\omega ^{2}\bar{a}^{5}-12\bar{a}+4\Lambda 
)^{2}}
\right] \,.  \label{aa3d}
\end{equation}

In the limit $\bar{a}\rightarrow 0$, Eq. (\ref{aa3d}) takes the form 
\begin{equation}
\frac{d^{2}\bar{a}}{dt^{2}}=\left( -\Lambda +\frac{3\epsilon 
^{2}}{16\Lambda 
}\right) \bar{a}^{-4}\,,  \label{asymp3D}
\end{equation}
cf. Eq. (\ref{asymp2D}). Equation (\ref{asymp3D}) predicts one property 
of
the 3D model correctly, viz., in the case $\Lambda <0$ and with a
sufficiently large amplitude of the ac component [$\epsilon >\left( 
4/\sqrt{
3}\right) \left| \Lambda \right| $, as it follows from Eq. 
(\ref{asymp3D}],
collapse takes place instead of spreading out. However, other results
following from the averaged equation (\ref{aa3d}) are {\em wrong}, as
compared to those following from the direct simulations of the full
variational equation (\ref{a3d}), which were displayed above, see Figs. 
2
and 3. In particular, a detailed analysis of the right-hand side of Eq. 
(\ref
{aa3d}) shows that it does not predict a stable FP for $\Lambda <0$, 
and
does predict it for $\Lambda >0$, exactly opposite to what was revealed 
by
the direct simulations. This failure of the averaging approach (in 
stark
contrast with the 2D case) may be explained by the existence of 
singular
points in Eqs. (\ref{delta_a3D}) and (\ref{aa3d}) (for both $\Lambda 
>0$ and 
$\Lambda <0$), at which the denominator $\omega 
^{2}\bar{a}^{5}-12\bar{a}
+4\Lambda $ vanishes. Note that, in the 2D case with $\Lambda >0$, for 
which
the stable state was found in the previous section [see Eq. 
(\ref{stable2D})], the corresponding equation (\ref{aa2d}) 
did not have singularities.

\subsection{Direct simulations of the Gross-Pitaevskii equation in the
three-dimensional case}

Verification of the above results given by VA against direct 
simulations of
the 3D version of the radial equation (\ref{nls}) is necessary. The 
partial
differential equation simulations were carried out by means of the 
method of lines implemented
with the DOPRI5 ODE solver and space discretization involving high 
order
finite differences, see the details in Appendix 2.
The relative error in the conservation of the number of atoms was 
limited by 
$10^{-8}$. In the absence
of the ac modulation, the energy was conserved with a relative error 
limited
by $10^{-8}$.

Quite naturally, in the case $\epsilon =0$ (no ac component) and 
$\Lambda <0$, the simulations show straightforward decay of the 
condensate (not
displayed here, as the picture is rather trivial). If an ac component 
of
sufficiently large amplitude is added, stabilization of the condensate
takes place temporarily, roughly the same way as is predicted by the
solution of the variational equation (\ref{a3d}). However, the 
stabilization
is not permanent: the condensate begins to develop small-amplitude
short-scale modulations around its center, and after about $50$ 
periods of the ac modulation, it collapses.

An example of this behavior is displayed in Fig. 4, for which 
$N=1,~\Lambda =-1$ and $\omega =10^{4}\cdot \pi $. Figure 4 shows 
radial
profiles of the density $|u(r)|^{2}$ at different instants of time.

%\begin{figure}[ht]
%\centerline{\psfig{figure=f05.ps,height=5cm,width=4.5cm,angle=-90}
%\psfig{figure=f06.ps,height=5cm,width=4.5cm,angle=-90}
%\psfig{figure=f07.ps,height=5cm,width=4.5cm,angle=-90}} 
%%was \label{f07}
%\caption{Time evolution of the condensate's shape $|u|^2(r)$ in the 
%presence
%of the strong and fast ac modulation ($\protect\omega= 
%10^4\dot\protect\pi, 
%\protect\epsilon=90$). From left to right the profiles of $u^2(r)$ 
%are shown at times $t=0.007, 0.01 $ and $0.015$.} 
%\label{Fig4}
%\end{figure}

Results presented in Fig. 4 turn out to be quite typical for the 3D 
case
with $\Lambda <0$. The eventual collapse which takes place in this case 
is a
nontrivial feature, as it occurs despite the fact that the dc part of 
the
nonlinearity drives the condensate towards spreading out. Therefore, a 
basic
characteristic of the system is a dependence of the minimum ac 
amplitude 
$\epsilon $, which gives rise to the collapse at fixed $\Lambda =-1$, 
versus
the ac frequency$\,\omega $. Several points marked by stars show this
dependence in Fig. 3. It is quite natural that the minimum value of 
$\epsilon $ necessary for the collapse grows with $\omega $. On the 
other
hand, for $\omega $ not too large, the minimum ac amplitude necessary 
for
the onset of collapse becomes small, as even a small $\epsilon $ is 
sufficient
to push the condensate into collapse during the relatively long 
half-period
when the sign of the net nonlinearity coefficient $\lambda (t)$ is 
positive,
see Eq. (\ref{omega}).

In the case of $\Lambda >0$ we have never been able to prevent the
collapse of the pulse. This is in agreement with
the analysis developed in the previous section on the basis of the
Hamiltonian of the averaged version of the GP equation, which showed 
that
the collapse cannot be stopped in the 3D case, provided that the 
amplitude
of the ac component is large enough. Besides that, this eventual result 
is
also in accordance with findings of direct simulations of the 
propagation of
localized 3D pulses in the above-mentioned model of the nonlinear 
optical
medium consisting of alternating layers with opposite signs of the Kerr
coefficient: on the contrary to the stable 2D spatial solitons 
\cite{towers},
the 3D spatiotemporal ``light bullets'' can never be stable in this 
model 
\cite{unpub}.

\section{Conclusion}

In this work, we have studied the dynamics of 2D and 3D Bose-Einstein
condensates in the case when the scattering length in the 
Gross-Pitaevskii
(GP) equation contains constant (dc) and time-variable (ac) parts. This 
may
be achieved in the experiment by means of a resonantly tuned ac 
magnetic
field. Using the variational approximation (VA), simulating the GP 
equation
directly, and applying the averaging procedure to the GP equation 
without
the use of VA, we have demonstrated that, in the 2D case, the ac 
component
of the nonlinearity makes it possible to maintain the condensate in a 
stable
self-confined state without external traps, which qualitatively agrees 
with
recent results reported for spatial solitons in nonlinear optics. In 
the 3D
case, VA also predicts a stable self-confined state of the condensate
without a trap, provided that the constant part of the nonlinearity
corresponds to repulsion between atoms. Direct simulations reveal that, 
in
this case, the stability of the self-confined condensate is limited in 
time.
Eventually, collapse takes place, despite the fact that the dc 
component of
the nonlinearity is repulsive. Thus, we conclude that the spatially 
uniform
ac magnetic field, resonantly tuned to affect the scattering length, 
may
readily play the role of an {\em effective trap} which confines the
condensate, and sometimes enforces its collapse. These predictions can 
be
verified in experiments.

\section*{Acknowledgements}

F.K.A. and B.A.M. appreciate hospitality of Instituto de Fisica Teorica 
--
UNESP (S\~{a}o Paulo, Brazil). The work is partially supported by 
FAPESP.
Authors are grateful to R.Galimzyanov for help with the numerical 
simulations.

\section{Appendix 1: calculation of the moments}

For the modulation analysis of section 3.2, we introduce the 
following integrals involving the Townes soliton.

The boundary value problem (\ref{BVP}) has been solved by discretzing using 
finite differences and using the shooting method. The solutions
give a residual smaller than $10^{-7}$.

The integrals have been calculated using the trapezoidal rule.
As a test the following integrals has been calculated, 
the norm $N(R_{s}) = 1.862... $ and hamiltonian
($H(R_{s}) =0$). 
For the other integrals we obtain

\begin{eqnarray}
I_{1} = \int_{0}^{\infty}r^{2}dr R_{T}^{2} = 1.7~~;~~\ 
I_{2} = \int_{0}^{\infty}rdr 
R_{T}^{2}(R_{T}^{\prime})^{2} = 2.529~~;\nonumber\\ 
I_{3} = \int_{0}^{\infty}rdr R_{T}^{4}(R_{T}^{\prime})^{2} = 5.730;\ 
~~
I_{4} = \int_{0}^{\infty} r^{2}dr R_{T}^{3}(R_{T}^{\prime})^3 = 
-3.109~~;~~\\
I_{5} = \int_{0}^{\infty}dr R^{4} = 11.472~~;~~ \ 
I_{6} = \int_{0}^{\infty}rdr R_{T}^6 = 11.312~~;~~ \ 
I_{8} = \int_{0}^{\infty} rdr R_{T}^{8} = 39.963, \nonumber\\
I_{9} = \int_{0}^{\infty} rdr R_{T}^{3}(R_{T}^{\prime})^{3} = -4.872~~;~~ \
I_{10} = \int_{0}^{\infty} rdr R_{T}^{3}(R_{T}^{\prime})^{2} = 3.669~~;~~ \ \nonumber\\
I_{11} = \int_{0}^{\infty} rdr R_{T}^{2}(R_{T}^{\prime})^{3} = -2.314.\nonumber
\end{eqnarray}

\section{Appendix 2: numerical procedure for solving the partial 
differential
equation} 

Following \cite{Fibich}, we have solved the cylindrical NLS equation
(\ref{nls}) using the method of lines where the solution is advanced
in time using an ordinary differential equation (ODE) solver and 
the spatial part is discretized using finite differences. Because
of its implicit character, this method allows for great stability 
and accuracy as well as giving the possibility of implementing 
directly the cylindrical Laplacian and its associated boundary 
conditions.

Specifically we use
as ODE solver the variable step Runge-Kutta of order 4-5 DOPRI5 
\cite{hairer}.
which enables to control the error made at each step and bound it by
a given tolerance. For all the runs presented the relative error 
at each step is below $10^{-8}$. The cylindrical Laplacian 
$\partial_r^2 +(D-1)\partial_r/r$ is approximated at each node $n$ of 
the grid
using the following formulas 
$$\psi_r|_n ={1 \over 12 
h}(\psi_{n-2}-8\psi_{n-1}+8\psi_{n+1}-\psi_{n+2})
+O(h^4) , $$
$$\psi_{rr}|_n ={1 \over 12 h^2}(-\psi_{n-2}+16\psi_{n-1}-30 \psi_n 
+16\psi_{n+1}-\psi_{n+2}) 
+O(h^4) , $$
where $h$ is the mesh size.
We have therefore a method to solve (\ref{nls}) that is $O(dt^4,h^4)$

The first node corresponds to $r=0$ and to its left we introduce two 
fictitious
points so that $\psi_r\_{r=0}=0$. At the right hand side boundary 
chosen
sufficiently far from the pulse, $\psi$ was set to be $0$, again in
two points. 

The number of mesh points was $4000$ and the tolerance of the 
integrator
set to $10^{-8}$. In all cases the ${\rm L^2}$ norm $N$ was conserved 
up
to $10^{-8}$ in relative error as was the Hamiltonian in the absence of
modulation. The latter quantity provided an accurate indicator of 
collapse.

\newpage
 
\begin{figure}
\caption{A typical example of the formation of a self-confined
condensate, revealed from direct simulations of Eq. (\ref{nls}) in the
two-dimensional case. The left panel shows pulse collapse in the 
absence
of modulation for $t\approx 0.3$. The right panel shows the modulated 
pulse with the same initial condition for $t\approx 0.6$.
The parameters are $\lambda_0=2.4,~\lambda_1=0.85$
$\omega = 100 \pi$ and $N=5$. }
\label{Fig1}
\end{figure}
\begin{figure}
\caption{The Poincar\'{e} section in the plane $(a,\dot{a})$ for 
$\Lambda=-1, \protect\epsilon=100, \protect\omega=10^4\dot\protect\pi 
$,
generated by the numerical solution of the variational equation 
(\ref{a3d})
with different initial conditions (see the text).}
\label{Fig2}
\end{figure}
\begin{figure}
\caption{The region in the $(\protect\epsilon, 
\protect\omega/\protect\pi)$
parameter plane where the numerical solution of Eq. (\ref{a3d}) with 
$\Lambda=-1$ predicts stable quasiperiodic solutions in the 3D case. 
Crosses
mark points where stable solutions were actually obtained. Stars 
correspond
to the minimum values of the ac-component's amplitude 
$\protect\epsilon$
eventually leading to collapse of the solution of the full partial
differential equation (\ref{nls}) with $\Lambda=-1$, see below.}
\label{Fig3}
\end{figure}
\begin{figure}
\caption{Time evolution of the condensate's shape $|u|^2(r)$ in the 
presence
of the strong and fast ac modulation ($\protect\omega= 
10^4\dot\protect\pi, 
\protect\epsilon=90$). From left to right the profiles of $u^2(r)$ 
are shown at times $t=0.007, 0.01 $ and $0.015$.}
\label{Fig4}
\end{figure}
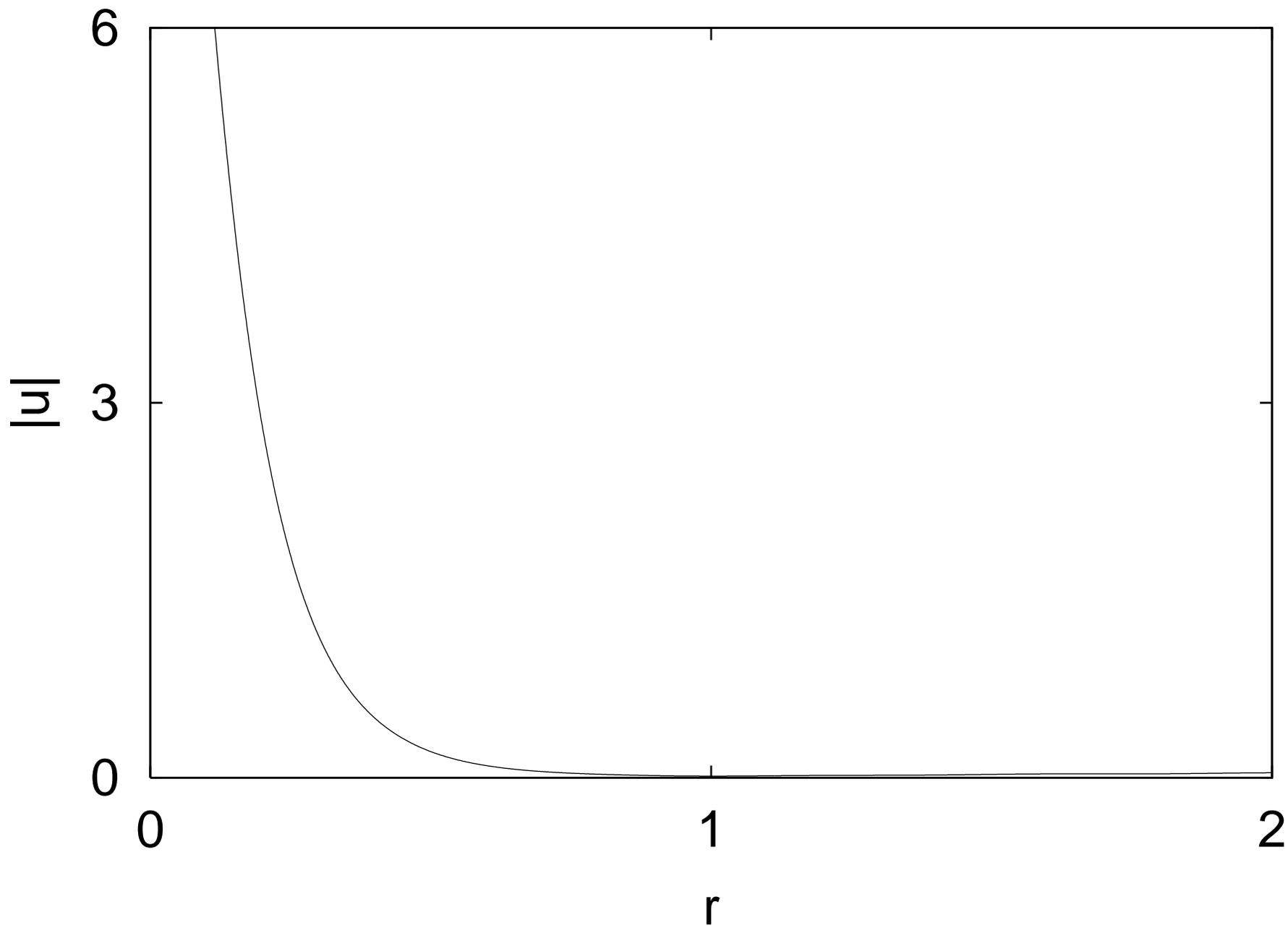
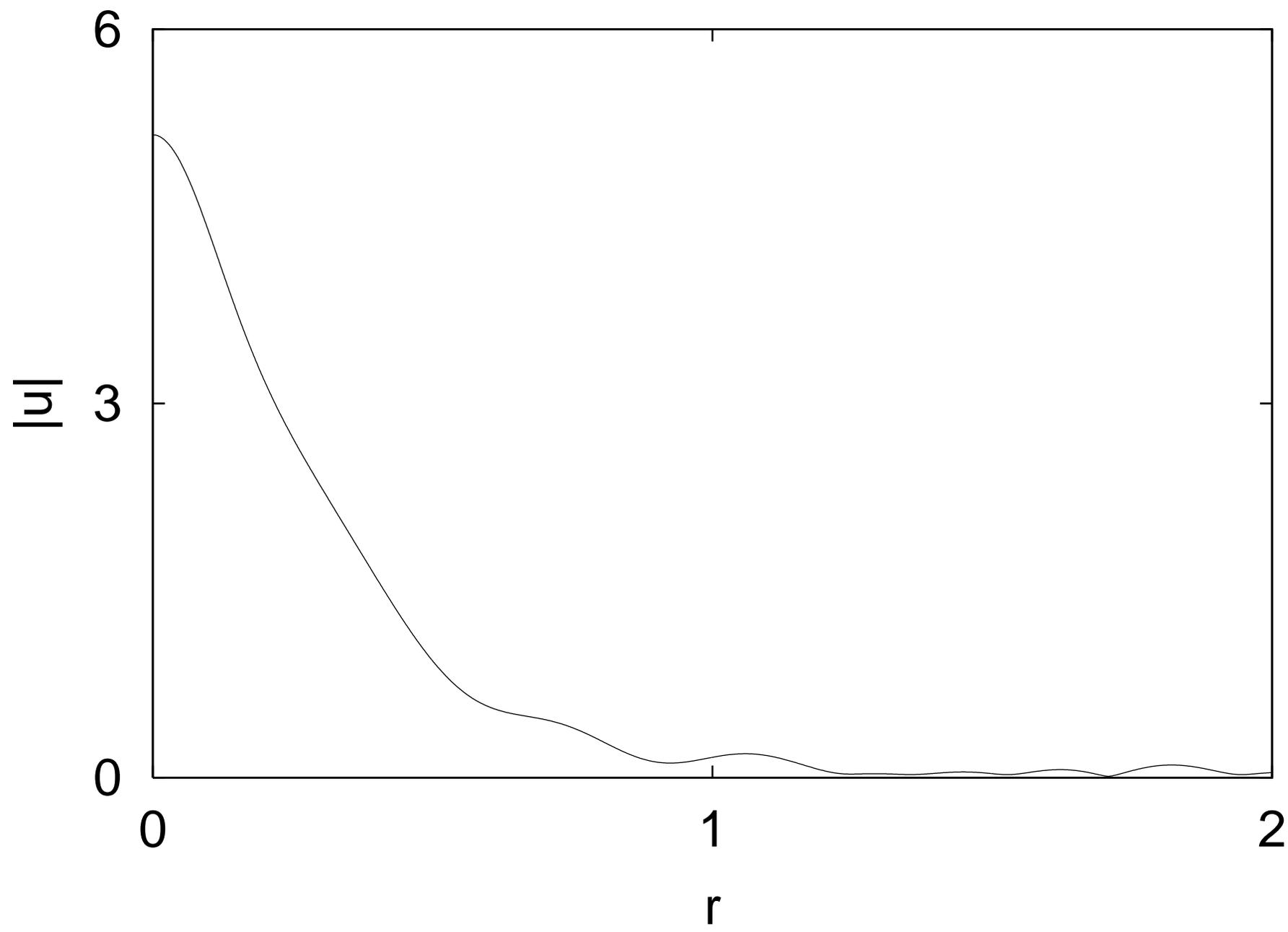
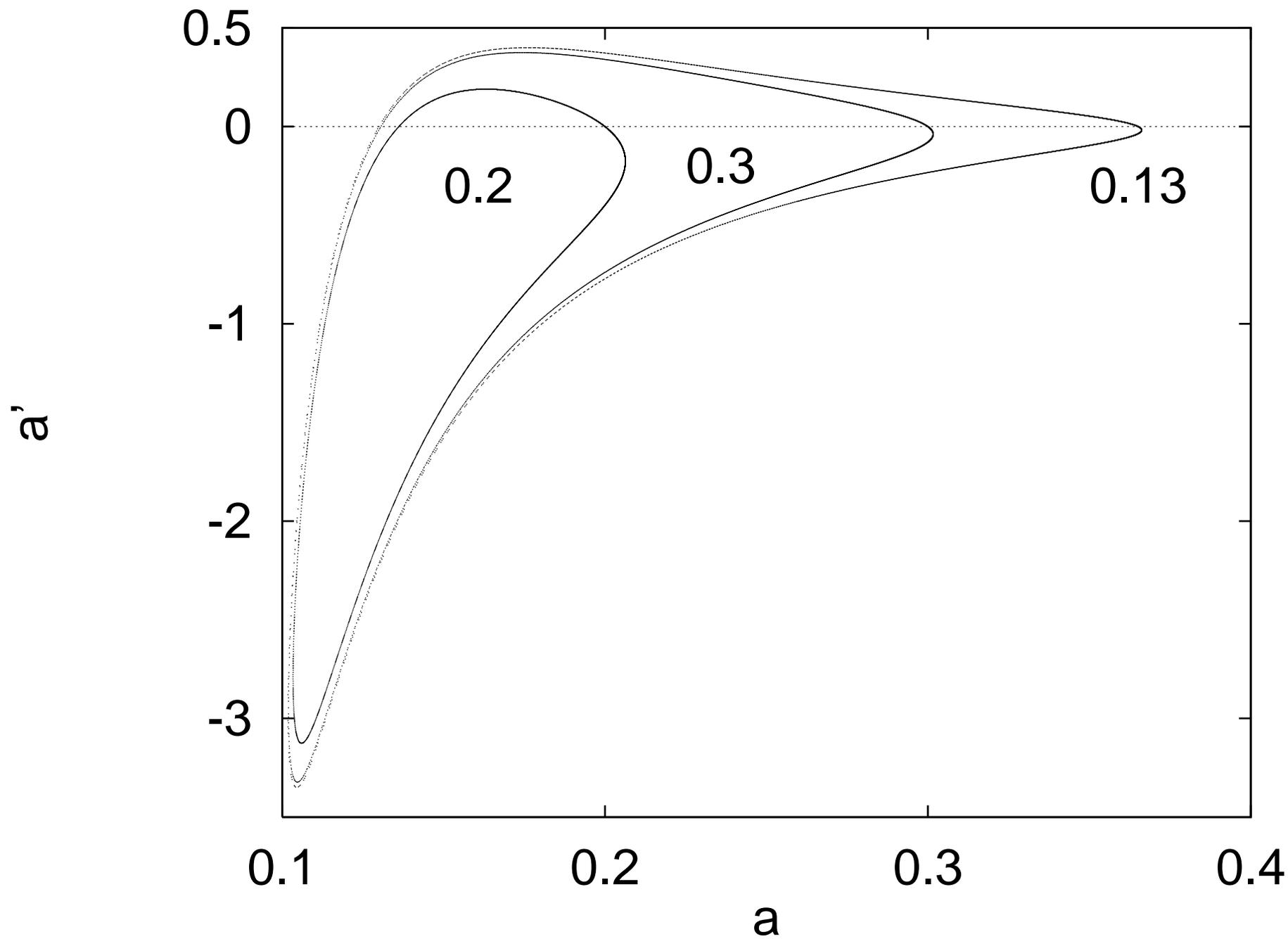
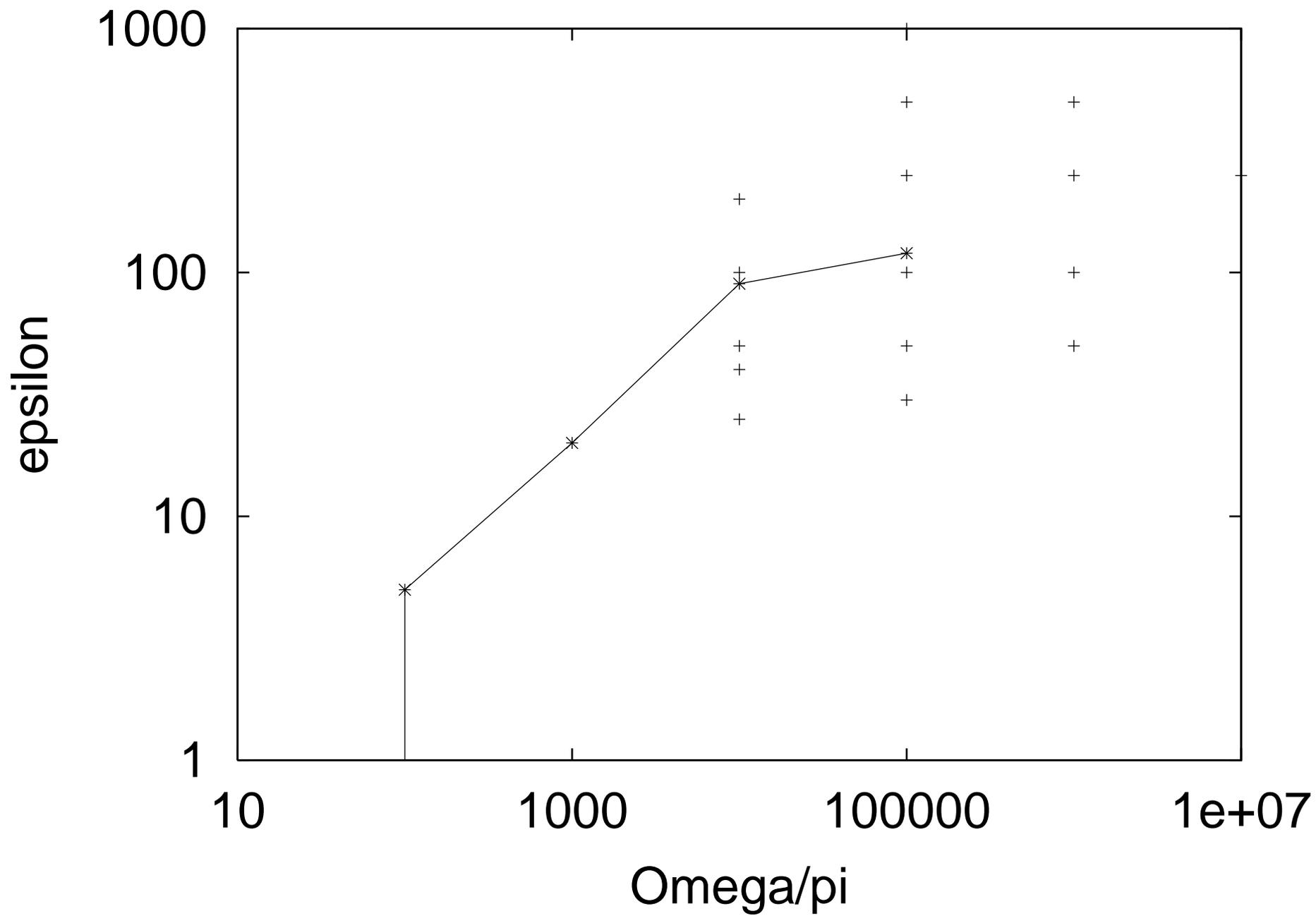
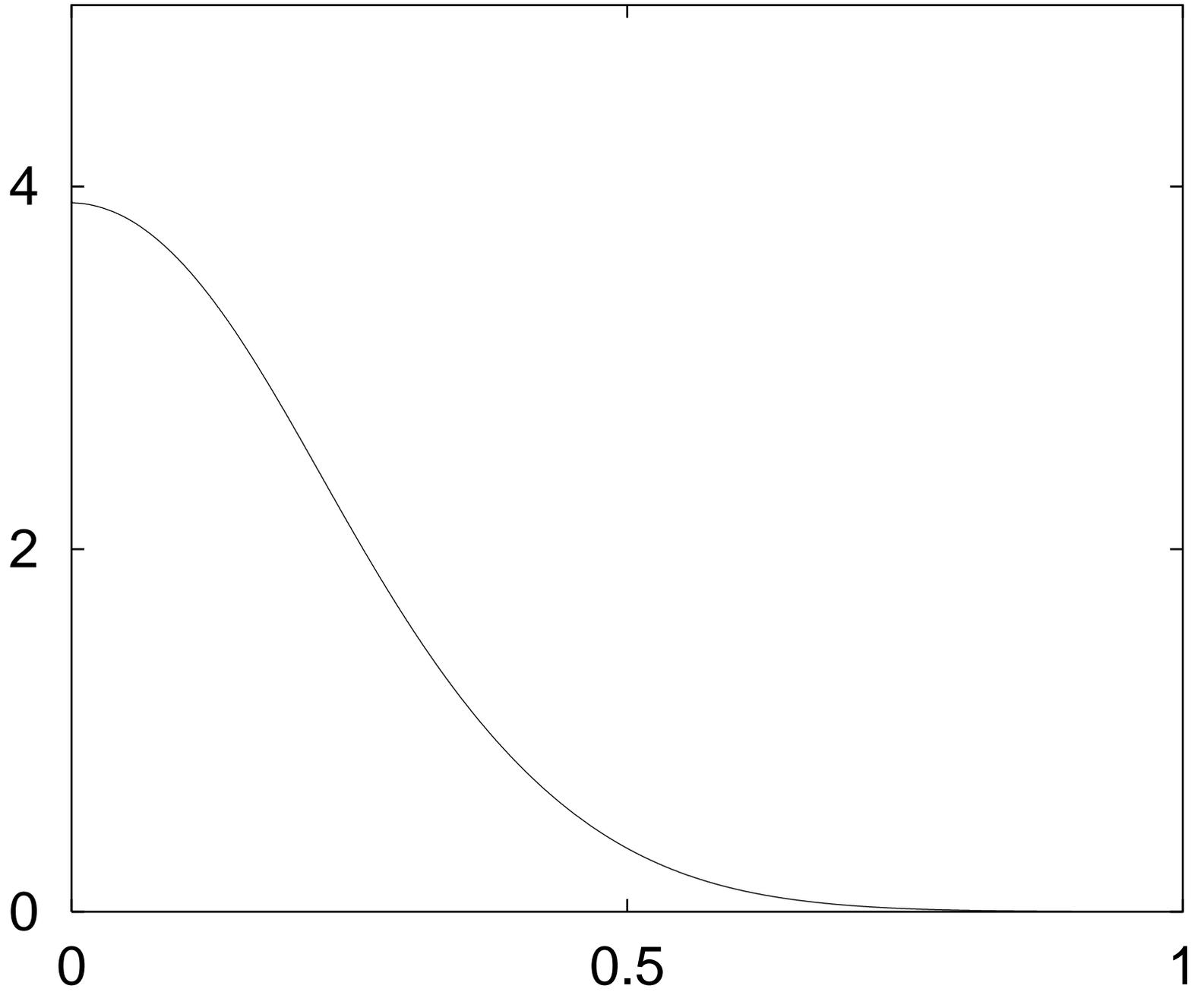
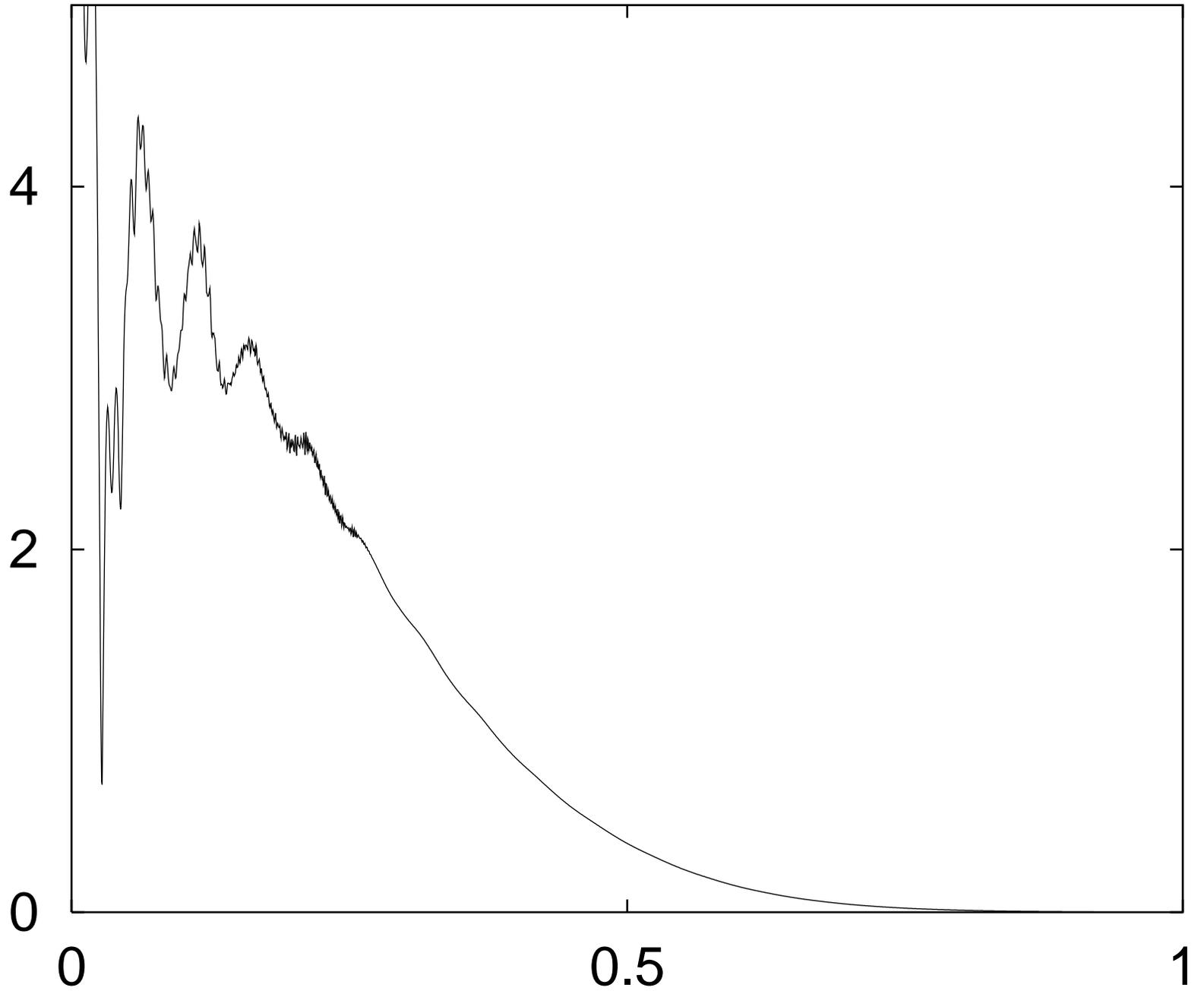
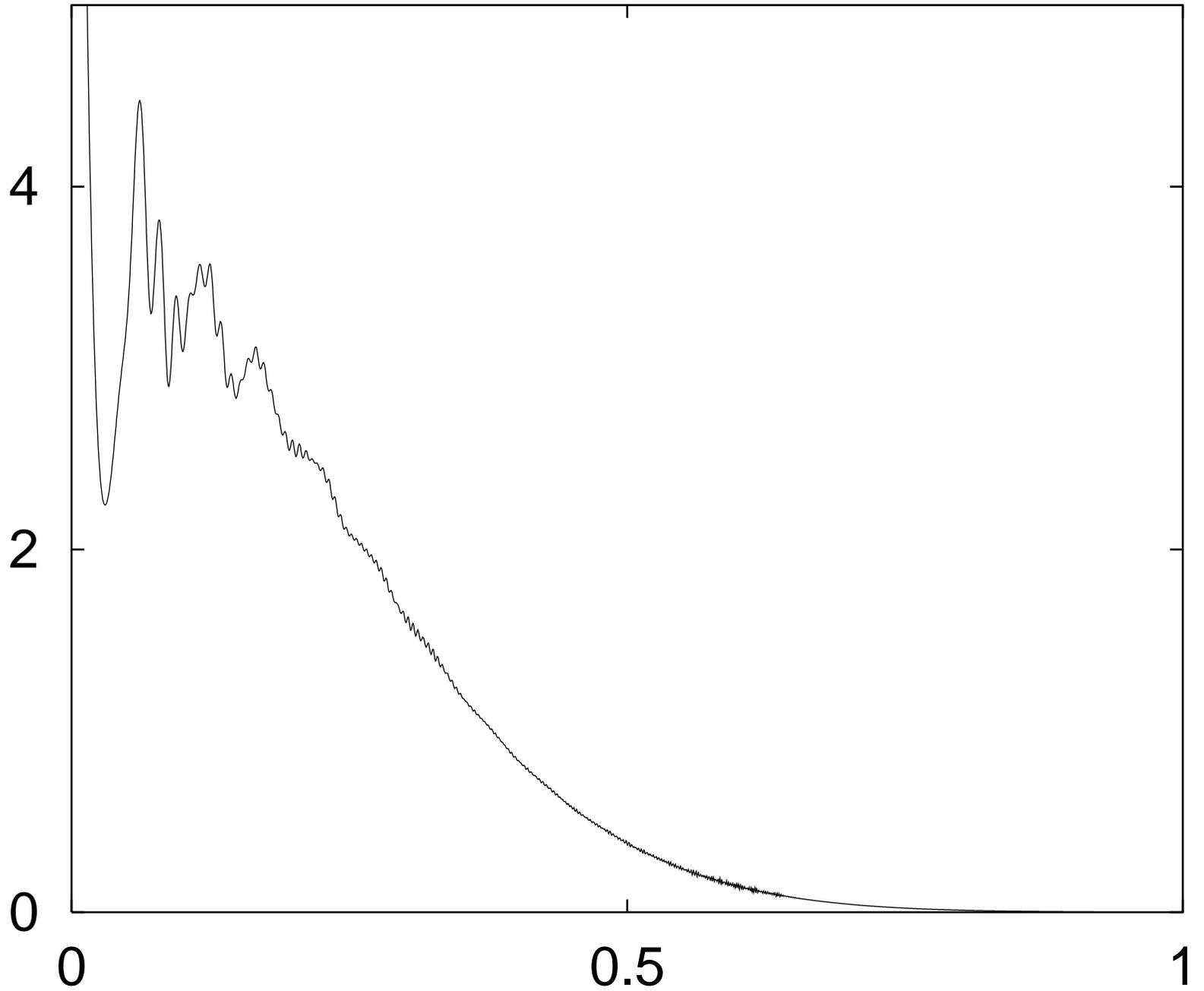
\end{document}